\documentclass{Interspeech}

\interspeechcameraready

\title{End-to-End DOA-Guided Speech Extraction in Noisy Multi-Talker Scenarios}

\author[affiliation={1\dagger}]{Kangqi}{Jing}
\author[affiliation={2\dagger}]{Wenbin}{Zhang}
\author[affiliation={2*}]{Yu}{Gao}

\affiliation{School of Information Science and Engineering}{Southeast University}{China}
\affiliation{AI Research Center}{Midea Group (Shanghai) Co.,Ltd.}{China}
\email{
\thanks{$\dagger$ Equal contribution.}
\thanks{* Corresponding author.}
jingkangqi@seu.edu.cn, \{zhangwb87, gaoyu11\}@midea.com}


\usepackage{comment}

\usepackage{cite}
\usepackage{amsmath,amssymb,amsfonts}
\usepackage{algorithmic}
\usepackage{graphicx}
\usepackage{textcomp}
\usepackage{xcolor}
\usepackage{multirow}
\usepackage{makecell}
\usepackage{threeparttable}
\usepackage{url}

\begin{document}

\maketitle

\begin{abstract}
    

    Target Speaker Extraction (TSE) plays a critical role in enhancing speech signals in noisy and multi-speaker environments. This paper presents an end-to-end TSE model that incorporates Direction of Arrival (DOA) and beamwidth embeddings to extract speech from a specified spatial region centered around the DOA. Our approach efficiently captures spatial and temporal features, enabling robust performance in highly complex scenarios with multiple simultaneous speakers. Experimental results demonstrate that the proposed model not only significantly enhances the target speech within the defined beamwidth but also effectively suppresses interference from other directions, producing a clear and isolated target voice. Furthermore, the model achieves remarkable improvements in downstream Automatic Speech Recognition (ASR) tasks, making it particularly suitable for real-world applications.

\end{abstract}

\section{Introduction}

In complex auditory environments with multiple sound sources, humans are able to selectively focus on a specific sound, a phenomenon commonly known as the ``cocktail party effect''. This ability allows us to attend to a target sound using various cues, such as its time-frequency pattern or the direction of arrival (DOA) \cite{haykin2005cocktail}. Target Sound Extraction (TSE) aims to replicate this selective attention mechanism by isolating and enhancing a desired sound source, typically human speech, from a noisy and reverberant background. This task is critical for applications in fields such as conference systems and hearing aids \cite{luo2023music, patterson2022distance}.
The primary challenges in TSE include background noise, speech interference, and reverberation, all of which can severely degrade the quality and intelligibility of the target speech \cite{wen2024neural}. Among these, speech interference, especially from competing speakers, poses the greatest difficulty. Speech recognition systems, in particular, struggle to distinguish the target speech from the interference, resulting in a significant drop in performance.

Recent advancements in Deep Neural Networks (DNNs) have made significant strides in overcoming the challenges of target sound extraction. These developments have proven especially effective in applications such as speech separation \cite{wang2023dasformer,wang2023tf}, enhancement \cite{lee2023deft}, and DOA estimation \cite{li2024robust}. In multichannel audio setups, DOA provides a crucial cue for locating the target speaker. This technique, referred to as Directional Speech Extraction (DSE), aims to isolate speech from a fixed \cite{kovalyov2023dsenet, tesch2022insights} or dynamically adjustable \cite{xu2020neural, tesch2023multi} spatial region. Recently, DSE approaches \cite{gu2024rezero, pandey2024all} have been further refined to leverage DOA information from multichannel mixtures, where the DOA cue is used to guide the extraction process. 

However, most existing models for TSE rely heavily on accurate DOA information to achieve optimal performance \cite{rascon2025direction}. When the DOA estimates are inaccurate or ambiguous, these models often fail to clearly define the spatial region for sound extraction, resulting in unpredictable outputs that may include both target speech and interfering noise. This limitation significantly degrades their effectiveness in real-world scenarios where precise DOA information is not always available.
The CDUNet model \cite{wen2024neural} proposing a directional TSE approach that integrates beamforming and DNNs for two-microphone and two-speaker setups. This model introduces the enhancement width as an input parameter to control the spatial region for speech enhancement. While innovative, CDUNet lacks generalization to more complex, multi-speaker environments.

In this work, we propose an end-to-end TSE model that leverages DOA embedding to extract speech within a selected beamwidth centered around the DOA. Our model introduces advanced voice focusing and zooming capabilities, enabling robust performance in scenarios involving multiple speakers. By efficiently extracting target speech signals from different directions, our approach significantly improves the performance of downstream ASR tasks. This makes it particularly suitable for real-world applications such as conference systems and smart assistants, where accurate and reliable speech extraction in highly overlapping multi-speaker environments is critical.
Code and audio examples are available at \url{https://github.com/jingkangqi/DSENet}.

\section{Method}
\begin{figure*}[t!]
  \centering
  \includegraphics[scale=0.7]{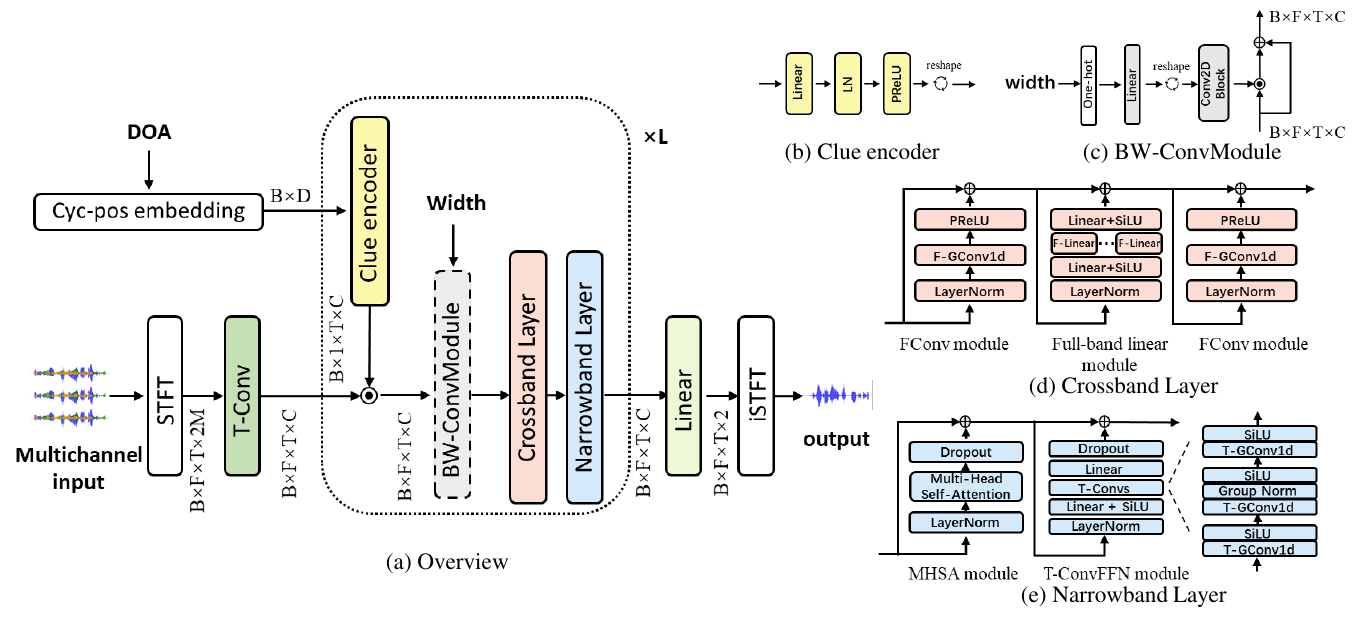}
  \caption{(a) The system overview of the proposed DSE model. The detailed structure of (b) Clue encoder, (c) BW-ConvModule, (d) Crossband Layer and (e) Narrowband Layer.}
  \label{fig:model_architecture}
\end{figure*}
\subsection{Problem formulation}

Let $\mathbf{X} \in \mathbb{R}^{M \times L}$ be the multichannel mixture recorded by an $M$-microphone array, where $L$ denotes the number of time samples. The signal at the $m$-th microphone can be expressed as:  
\begin{equation}
    \mathbf{X}_m(t) = \sum_{i=1}^{N} {s}_i(t) * \mathbf{H}_i^m(t) + \mathbf{N}_m(t),
\end{equation}
where ${s}_i(t) \in \mathbb{R}^{ L} $ represents the clean speech signal of the $i$-th speaker, $\mathbf{H}_i^m(t)$ denotes the room impulse response (RIR) from the $i$-th speaker to the $m$-th microphone, and $\mathbf{N}_m(t)$ represents the background noise at the $m$-th microphone. 

Given a target DOA \( \theta_{\text{target}} \) and a beamwidth \( \theta_{\text{beam}} \), our model aims to extract speech signals ${s}_c $ exists in \( \theta_{\text{beam}}=[\theta_{\text{target}}-\theta_{\text{width}}, \theta_{\text{target}}+\theta_{\text{width}}] \) , while suppressing interference from other directions outside the beam. The selected beamwidth \( \theta_{\text{beam}} \) may contain more than one speaker, in which case the model will retain and output all voices within this range. If no speech is present in the range, the model outputs a signal that is either close to silence or silent.

\subsection{Model Architecture}
The overview of the proposed architecture is depicted in Figure~\ref{fig:model_architecture}. First, a multichannel input is converted into a 2M × T × F complex spectrogram by short-time Fourier transform (STFT), where M denotes the number of microphone, T and F denote the number of time and frequency bins, respectively. 
The spectrogram is then processed by a convolutional input layer, expanding the channel dimension to C while encoding spatial features and local time-frequency information. 
The feature sequence is then processed through L stacked layers. In each layer, the clue encoder refines the DOA-related features for spatial selectivity and target extraction. The BW-ConvModule limits the beamwidth around the DOA, dynamically restricting the spatial region to improve focus on target speech. The Crossband blocks capture spectral dependencies by modeling frequency-wise correlations, while the Narrowband blocks focus on temporal patterns within each frequency band.
The final layer's output is projected to target STFT coefficients via linear transformation, with iSTFT reconstructing the enhanced speech signal $\hat{s}_c$.

\subsubsection{DOA-based clue embedding and beamwidth control}

While the one-hot vector provides a unique representation for each direction, it fails to capture the inherent periodicity, resulting in an abrupt transition from \( 359^\circ \) to \( 0^\circ \). Moreover, the one-hot vector requires a high-dimensional encoding (360 dimensions), which is computationally inefficient. To address these issues, we adopt cyclic positional (cyc-pos) encoding \cite{lee2023spatio}, which effectively preserves the continuity of direction while reducing the dimensionality. The cyc-pos vector $\mathbf{PE}_{\text{cyc-pos}}(\phi) \in \mathbb{R}^D$ for an embedding dimension of $D$ is defined as:  
\begin{equation}
\begin{aligned}
\mathbf{PE}_{\text{cyc-pos}}(\phi, 2j) &= \sin\left(\sin(\phi) \cdot \frac{\alpha}{10000^{2j/D}}\right) \\
\mathbf{PE}_{\text{cyc-pos}}(\phi, 2j + 1) &= \sin\left(\cos(\phi) \cdot \frac{\alpha}{10000^{2j/D}}\right)
\end{aligned}
\end{equation}
where $j \in \left[0, \frac{D}{2}\right)$, and $\alpha$ is a scaling factor.

The DOA embedding is then broadcast along the time dimension and processed through the clue encoder, as illustrated in Figure~\ref{fig:model_architecture}(b). The clue encoder comprises a linear layer, followed by layer normalization (LN) \cite{lei2016layer} and parametric rectified linear unit (PReLU) \cite{he2015delving}. The encoded DOA embedding is applied via element-wise multiplication with the output of the T-Conv and Narrowband layers excluding the final layer. 

The BW-ConvModule shown in Figure~\ref{fig:model_architecture}(c), dynamically adjusts the beamwidth by processing a width input through a one-hot encoding, a Linear layer, and a 1×1 Conv2D to generate a mask. This mask filters out noise outside the desired beamwidth, while residual connections preserve essential information for robust target speech extraction.

\subsubsection{Crossband and Narrowband layers}

The Crossband and Narrowband layers have the same structure as that described in \cite{quan2024spatialnet}, designed to learn complex spatial information, as shown in Figure~\ref{fig:model_architecture}(d)(e).


The Crossband layer comprises two frequency-convolutional modules and a full-band linear module, processing each time frame independently. The F-GConv1d module utilizes grouped frequency convolutions to model local spectral dependencies. The full-band linear module first compresses channels ($C \rightarrow C'$), followed by a set of frequency-wise linear layers. To improve parameter efficiency, the same F-Linear networks are shared across all instances of the Crossband layer. Finally, a linear layer restores the channel dimension to $C$.

The Narrowband layer captures temporal dependencies by processing each frequency independently. The multihead self-attention (MHSA)\cite{vaswani2017attention} module computes spatial similarities within each frequency, facilitating the separation of speech components originating from different directions. 
The time-convolutional feedforward network (T-ConvFFN) enhances temporal modeling via a sequence of operations: a linear layer expands the hidden dimension from $C$ to $C''$, followed by grouped 1-D convolutions along the time axis with group normalization \cite{wu2018group}, before a final linear layer restores the original dimensionality $C$.

The Crossband and Narrowband layers are interleaved to enhance the model's ability to differentiate and extract target signals while effectively noise, reverberation, and interference.

\subsection{Loss Function}
The proposed model is trained using a combination of a spectral magnitude loss and a scale-invariant signal-to-distortion ratio (SI-SDR) \cite{8683855} loss. The overall loss function is formulated as:
\begin{equation}
    \mathcal{L} = \mathcal{L}_{\text{Mag}} + \lambda \mathcal{L}_{\text{SI-SDR}},
\end{equation}
where \( \lambda \) is a weighting factor. The spectral magnitude loss, similar to \cite{wang2023tf}, enforces consistency in the TF domain and is defined as:
\begin{equation}
    \mathcal{L}_{\mathrm{Mag}} = \frac{\left\lVert |\mathrm{STFT}(\hat{s}_c)| - |\mathrm{STFT}(s_c)| \right\rVert_1}{\left\lVert |\mathrm{STFT}(s_c)| \right\rVert_1}.
\end{equation}
To improve the perceptual quality of the extracted speech, we incorporate the SI-SDR loss:
\begin{equation}
    \mathcal{L}_{\text{SI-SDR}} = - \sum_{c=1}^{C} 10 \log_{10} \left( \frac{\left\lVert s_c \right\rVert_2^2}{\left\lVert \hat{s}_c - \alpha_c s_c \right\rVert_2^2} \right),
\end{equation}
where \( \alpha_c \) is the optimal scaling factor given by
\begin{equation}
    \alpha_c = \frac{s_c^T \hat{s}_c}{s_c^T s_c}.
\end{equation}

\section{Experiment}
\subsection{Dataset}

For this study, the speech data is sourced from the LibriSpeech corpus \cite{panayotov2015librispeech}, while the background noise is taken from the DEMAND dataset \cite{thiemann2013diverse}. The room acoustics, including reverberation and microphone array configuration, are simulated using the \verb|pyroomacoustics|\footnote{\url{https://github.com/LCAV/pyroomacoustics}} \cite{scheibler2018pyroomacoustics} package. The room's width and depth are randomly sampled from the range [6, 9] m, with a fixed height of 3 m. The reverberation time (RT60) of the room is randomly varied within the range [0.3, 0.5] s. A 3-channel circular microphone array with a radius of 30 mm is placed at the center of each room, at a height of 1 m from the floor.

Each audio mixture contains speech from six simultaneous speakers combined with an independent noise source. All sources are randomly placed in the room, at least 0.3 m away from the walls. The Root-Mean-Square (RMS) level of each mixed signal is randomly sampled from [-20, -15] dB. 
All audio mixtures are 4 second-long, with a sampling rate of 16 kHz. The training, validation, and test sets contain 14.4k, 3.6k, and 2k mixtures, respectively. 

\subsection{Implementation Details}
 The network consists of \( L = 8 \) blocks, with hidden units configured as \( C = 192 \), \( C' = 8 \), and \( C'' = 192 \). The kernel sizes of TConv1d, T-GConv1d, and F-GConv1d are set to 5, 5, and 3, respectively, with all group numbers set to 8. 

The input STFT uses a window length of 16 ms (256 samples) and a hop length of 8 ms (128 samples), employing a Hanning window and 129 FFT frequency bins. Training batch size is 4, using the Adam \cite{diederik2014adam} optimizer with an initial learning rate of 0.001, which decays exponentially by a factor of 0.99 per epoch. The training is conducted in two stages:

\begin{itemize}
     \item \textbf{Stage 1: Target Extraction (100 epochs)} 
     Using DOA inputs from active speech sources, with the BW-ConvModule \textbf{disabled}. The SI-SDR loss weight is set to \(\lambda = 0.5\).

     \item \textbf{Stage 2: Beam Adaptation (50 epochs)} 
     The beamwidth input to BW-ConvModule is selected from one of \(15^{\circ}\), \(30^{\circ}\), or \(45^{\circ}\). For each selection, 90\% DOA inputs are active (\( \theta_{\text{beam}} \) containing speech sources), while the remaining 10\% are inactive (\( \theta_{\text{beam}} \) without speech sources).
     In inactive cases, the model generates a small, fixed signal instead of complete silence or noise. Silent periods contribute no meaningful information to SI-SDR calculation, and may cause training instability or large loss fluctuations. Here, we use a 20Hz reference tone ($-$60 dB RMS).
     The SI-SDR loss weight is decayed to $\lambda = 0.05$.

\end{itemize}

\section{Results and Analysis}
\subsection{Ablation Study and Comparison results}

To evaluate our proposed method's effectiveness, we compute three metrics: SDR improvement (SDRi)\cite{vincent2006performance}, SI-SDR improvement (SI-SDRi), and perceptual evaluation of speech quality (PESQ)\cite{rix2001perceptual}. We also analyzed the parameter count (Para.) to assess computational efficiency. For evaluation fairness and simplicity, we use the model trained int Stage 1, where the BW-ConvModule is disabled, and set the input DOA such that only one speaker falls within the beam width.

\vspace{-0.3cm}
\begin{table}[h!]
\centering
\setlength{\abovecaptionskip}{0.1cm} 
\setlength{\belowcaptionskip}{-0.1cm} 
\caption{Performance comparison of DSE Models and clues}
\label{table:results}
\setlength{\tabcolsep}{1mm}{
\begin{tabular}{|c|c|c|c|c|c|c|c|} 
\hline
\multirow{2}{*}{Model}    & \multirow{2}{*}{Emb.} & \multirow{2}{*}{D}  & \multirow{2}{*}{\(\alpha\)} & {\multirow{2}{*}{Para.}} & \multicolumn{3}{c|}{Evaluation Metrics$\uparrow$}         \\ 
\cline{6-8}
&   &   &   & \multicolumn{1}{l|}{}  & SI-SDRi   & SDRi   & PESQ    \\ 
\hline
\multicolumn{5}{|c|}{Noisy}   & -17.30    & -9.56       & 1.09     \\ 
\hline
MVDR   & -    & -    & -   & -   & -1.62     & -1.27     & 1.09   \\ 
\hline
JNF   & one-hot   & 360    & -    & 1.34M   & 10.22   & 8.04   & 1.10  \\ 
\hline
\multirow{4}{*}{Proposed}  & one-hot   & 360   & -   & 1.89M   & 16.21  
& 12.39  & 1.31     \\ 
\cline{2-8}
& \multirow{3}{*}{cyc-pos}   & \multirow{3}{*}{40} & 10   & \multirow{3}{*}{1.40M}   & 16.60          & 12.31          & 1.29           \\
&     &     & 20   &    & \textbf{18.29} & \textbf{13.99} & \textbf{1.40}  \\
&     &     & 40   &    & 14.82          & 11.19          & 1.25           \\
\hline
\end{tabular}}
\end{table}

\vspace{-0.3cm}
We conducted an ablation study to evaluate the impact of different DOA embedding types. The results in Table~\ref{table:results} demonstrate that cyclic positional encoding vectors outperform one-hot encoding. Specifically, the one-hot encoding achieves an SI-SDRi of 16.21 dB with 1.89M parameters, while the best-performing cyc-pos configuration attains an SI-SDRi of 18.29 dB with only 1.40M parameters.
The smooth and periodic nature of cyc-pos vectors across angles facilitates better integration into spatial feature representations, while simultaneously reducing the parameter count. However, when the scaling factor \(\alpha\) is set too large, performance degrades due to the loss of smoothness and the introduction of rapid variations.

For comparison, we selected the minimum variance distortionless response (MVDR) filter \cite{affes1997signal} and the Joint Spatial and TempoSpectral Non-linear Filter (JNF) \cite{tesch2023spatially} as baseline models. JNF is a deep neural network (DNN)-based spatially selective filter (SSF) that leverages a recurrent neural network (RNN) layer initialized with the target direction to spatially steer and extract the speaker of interest.
Due to the high complexity of the scenario involving up to six highly overlapping speakers, the MVDR method, which serves as a conventional benchmark, fails to effectively handle such an extremely challenging environment, resulting in degraded performance, which highlights the difficulty of the task.
JNF, which can be attributed to a better use of the spatial information achieve advanced results in Table~I. However, with cyc-pos vectors configured at \(D = 40\) and \(\alpha = 20\), our model significantly outperforms JNF, achieving higher SNRi, SI-SNRi, and PESQ scores while maintaining a comparable parameter count.

\subsection{Gain-pattern Analysis}
\begin{figure}[b!]
  \centering
  \includegraphics[width=\linewidth]{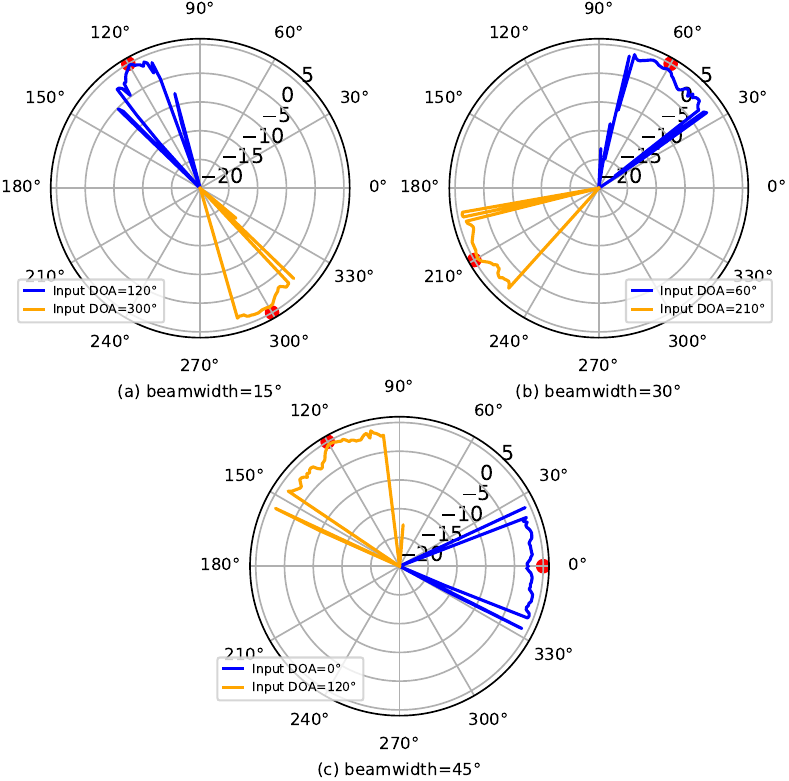}
  \caption{Gain-pattern analysis for different input beamwidths: (a) 15°, (b) 30°, and (c) 45°. Two random DOA inputs are tested for each beamwidth, with the input DOA indicated by a red dot.}
  \label{gain pattern}
  \renewcommand{\floatpagefraction}{.9}
\end{figure}
To facilitate the understanding of the complex spatial patterns incorporates reverberation effects, the methodology proposed in \cite{shmaryahu2022importance} introduces a simplified 1D representation of the array responses. In this section, we further explore the spatial properties of our method by plotting the gain pattern. 

The gain pattern is empirically generated by moving a speaker along a circular trajectory around the fixed microphone array, evaluating the model's response to the reverberant signal as a function of DOA.
Let the input and output signals be denoted by \textbf{x(t)} and \(y(t)\), respectively. The power of the dry speech component of \textbf{x(t)} in reference channal is defined as \(P_{\text{in}} = \mathbb{E} \left[ | x_{\text{dry}}(t) |^2 \right]\), where \(\mathbb{E}[\cdot]\) denotes the expected value. The power of the output signal is \(P_{\text{out}} = \mathbb{E} \left[ | y(t) |^2 \right]\). The gain in a given direction is computed as:

\begin{equation}
Gain \left( \text{dB} \right) = 10 \log_{10} \left( \frac{P_{\text{out}}}{P_{\text{in}}} \right)
\end{equation}

We compute the gain for each direction and plot the resulting gain map in Figure~\ref{gain pattern}. The model significantly enhances the target speech signal within the selected beamwidth centered around the input DOA, with the gain in the target region being substantially higher while the gain from other directions remains below $-20$ dB, effectively suppressing interference sources. The results demonstrate the model’s capacity to simultaneously utilize both DOA and beamwidth inputs.
Notably, a spillover of approximately $5^{\circ}$-$10^{\circ}$ is observed, which is more pronounced in wider beamwidths (e.g., $45^{\circ}$) compared to narrower ones (e.g., $15^{\circ}$). This spillover is an acceptable trade-off given the model's ability to maintain effective speech extraction within the target region.
The beamwidth explicitly defines the speech extraction range and offers a tolerance for DOA estimation errors, making it more aligned with practical requirements. 

\subsection{Effective Speaker Extraction Capability}
\begin{figure}[h!]
  \centering
  \includegraphics[width=\linewidth]{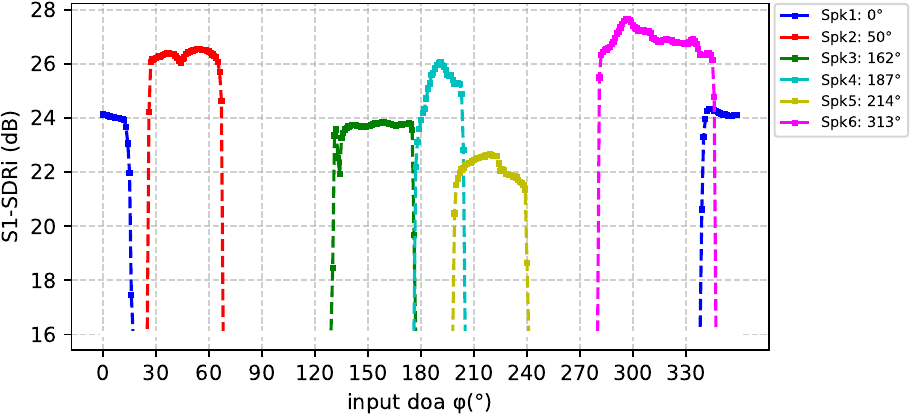}
  \caption{SI-SDRi for six speakers under various input DOAs with a fixed beamwidth of 30°, The DOAs of the six speakers (Spk1-Spk6) are: 0°, 50°, 162°, 187°, 214°, and 313°.}
  \label{fig:gain pattern}
\end{figure}

This work primarily evaluates the model's ability to extract clear speech signals corresponding to different speakers under varying input DOAs. To this end, we calculate the SI-SDRi for each speaker based on the model's output, with the beamwidth input fixed at 30°. As shown in Figure~\ref{fig:gain pattern}, when a speaker's DOA falls within the beamwidth, the corresponding SI-SDRi is substantially improved. As the input DOA varies, the model adaptively extracts speech from different speakers while effectively suppressing interference from other directions.

\subsection{Downstream ASR Performance}

For evaluating ASR performance, we additionally generated two datasets with different numbers of speakers, each comprising 1,000 samples. During testing, the input DOA was adjusted to correspond to the range of active sound sources.

\begin{table}[h!]
\centering
\setlength{\abovecaptionskip}{0.1cm} 
\setlength{\belowcaptionskip}{-0.1cm} 
\caption{WER (\%) results on different dataset }
\label{tab:asr_performance}
\begin{tabular}{c|ccc} 
\toprule
Mixed & Noisy & JNF & Proposed     \\ 
\hline
2spk   & 82.10    & 38.06  & \textbf{10.52}  \\ 
3spk   & 96.52    & 57.04  & \textbf{17.31}  \\ 
\bottomrule
\end{tabular}
\end{table}
\vspace{-0.2cm}

Table~\ref{tab:asr_performance} shows the Word Error Rate (WER) results. Our model achieves significantly lower WER compared to the JNF method, particularly as the number of speakers increases. This performance advantage underscores the model's potential for real-world applications involving multiple speakers, such as smart home devices, where it can accurately capture user commands amidst background noise.

\section{Conclusion}

In this work, we propose a novel end-to-end target speaker extraction model that leverages DOA and beamwidth as soft constraints to form an adaptive neural beam, dynamically focusing on the target speech even in highly complex multi-speaker environments. By integrating DOA and beamwidth embeddings, our approach efficiently captures spatial and temporal features, enabling robust performance in scenarios with significant speaker overlap and background noise. In future work, we will explore the deployment of this model on edge devices to further enhance its practicality in real-world scenarios.

\bibliographystyle{IEEEtran}
\bibliography{mybib}

\end{document}